# Conditions for Existence of Compact Stationary Relativistic Beam Torus

L A Sukhanova and Yu A Khlestkov

*Moscow Engineering Physics Institute (State University), Moscow, Kashirskoye shosse, 31, Russian Federation*

E-mail: khlestkov@yandex.ru

**Abstract**
Conditions for obtaining the absolute magnetic trap (AMT) to be used for the controlled fusion by means of high power relativistic charged beam have been considered. It is shown that AMT can not be created by external electromagnetic fields. Also the virial theorem implies that strict stationary state of the system does not exist in empty flat Minkowski space-time in case of positive beam pressure and positive electromagnetic field energy density.

## 1. Introduction

This work has been performed with the purpose of study the possibilities of charged beams physics, differential geometry and topology, used in general relativity theory, for solving the controlled fusion actual problems.

The high-temperature plasma confinement in a limited region by means of magnetic field pressure is one way to obtain a fusion reactor for the fusion energetic station. Magnetic field pressure prevents plasma contact with reactor walls. In this problem toroidal magnetic traps (tokamak concept) seem to be more perspective.

However, plasma has finitesimal life time in existing experimental research installations. This life time is limited by instabilities development. In energetic installations it has to be not less than the time necessary for the achievement of positive energy output when the generator energy produced by means of fusion reaction is much higher than the energy losses for plasma heating, plasma confinement and thermal and other radiation losses.

It is possible that the problem is that the magnetic trap formed by external sources currents and plasma own currents combination is not absolute. Confinement field formation by charged media own currents without external sources is a possible way to raise magnetic plasma confinement effectiveness. It can be realized, for example, by means of high power relativistic beam (HPRB) of electrons, protons or ions injection into tokamak. In contrast to external sources such HPRB[1] can produce confinement field without instabilities. It allows to raise plasma confinement time.

Such a problem statement is considered by the various authors, e.g. [1,2]. But to contrast with the these works we consider exact relativistic task without any approaches. This work continues also the known works of Shafranov et. al. [3,4] about plasma balance in toroidal traps. It is shown in these works that the plasma balance in its own longitudinal (toroidal) magnetic field does not exist and the confinement fields formed by means of external current windings is necessary for stationary state obtaining [3].

The extension of charged medium compact states in its own and external electromagnetic fields investigation in this work consists in the next problems. At first the unrestricted relativistic case of exact equations of motion and Maxwell's equations and unrestricted confinement fields combination as well as continuous closed magnetic surface without singularities (absolute magnetic trap, AMT) in hydrodynamic model is considered. In the previous investigation, however, the problem has been considered in non-relativistic approximation of quasistationary electromagnetic field with small velocities of charged particles motion in comparison with light speed.

---

1 In a high power charged beam with current much more than the so-called alphven current $m_0 c^3/e$ ($m_0$, $c$, $e$ – the fundamental rest mass, velocity of light and electric charge) particles movement is determined by its nonlinear interaction with its own electromagnetic field. The high power charged beam is necessary relativistic i.e. its Lorentz factor $\gamma \gg 1$ [10].

This work is devoted only to beam aspect of absolute magnetic trap (AMT) formation. HPRB interaction with high temperature plasma and plasma problems of heating, confinement are not considered.

## 2. Absolute magnetic trap existence condition

What are the general conditions of AMT existence? Does stationary compact state of charged medium in its own electromagnetic field exist? What are the conditions of its existence? Consider the first problem at first.

Statement 1. The absolute magnetic trap exists in the torus geometry of the charged medium own field.

1. The stationary nonradiative state of insular compact charged medium is axially-symmetrical [5].

2. AMT is an axially-symmetrical closed continuous magnetic surface without singularities. The magnetic field is tangent to this surface.

3. Only manifolds with zero Euler characteristic has continuous tangent vectors field [6].

4. In 3-dimensional space Euler characteristic equals zero only in torus topology [7].
So AMT should be an axially-symmetrical embedded magnetic toroids set created by electric charge currents.

5. Because an ideal current source has singularity (it has breakage because external coils have two disconnected endings) the magnetic surface produced with its participation has local heterogeneity. Therefore the magnetic field of external source can not form AMT.

6. Because an integral set of charged particles vortical trajectories in the compact beam torus (CBT) is continuous its magnetic field does not have singularities. Therefore in this case the necessary condition of AMT existence is realized.

So the necessary condition of AMT existence is the formation of relativistic charged medium compact toroidal state (HPRB) in its own electromagnetic field.

## 3. CBT stationarity condition : conservation laws implications

Let $T^{\mu\nu}$ is the energy momentum tensor of charged medium in its own and external electromagnetic fields ($\mu, \nu = 0,1,2,3$),

$$T^{\mu\nu}_{;\nu} = 0 \qquad (1)$$

– equations of motion [8].

A stationary compact system with flat asymptotic form has two isometries. When equations of motion are carried out two Killing vectors (mobility directions) $\xi^\mu_A$ ($A=0,3$), conform these isometries. The first one $\xi^\mu_0$ is the time-like, the second one $\xi^\mu_3$ is the space-like, cyclic one. This implies that the frame of reference with universal time exists in the stationary case. In this frame of reference $\xi^\mu_A = \delta^\mu_A$, $g_{Aa} = 0$ ($a=1,2$) and it is axially-symmetrical [5].

Multiplying $T^{\mu\nu}$ and $\xi^\mu_A$, taking into account Killing equations $\xi^\mu_{A(\mu;\nu)} = 0$, after integration of (1) over 4-manifold volume $\Omega^{(4)}$, and use Gauss theorem [7] we have from (1) :

$$0 = \int_{\Omega^{(4)}} \left(T^{\mu\nu}\xi_{\nu A}\right)_{;\mu} d\Omega = \oint_{\Sigma^{(3)}=\partial\Omega} T^{\mu\nu}\xi_{\nu A} d\Sigma_\mu ,$$

from which two integrals of motion existence follows :

$$K_A = \int_{\Sigma^{(3)}} T_\nu^\mu \xi_A^\nu d\Sigma_\mu = const. \qquad (2)$$

Here $\Sigma^{(3)}$ is 3-hypersurface limiting 4-volume $\Omega^{(4)}$.

On the other hand according to Komar [5] ( $\kappa = \dfrac{8\pi k}{c^4}$ is Einstein constant) :

$$\begin{aligned}K_A &= -\frac{2}{\kappa}\oint_{\Sigma^{(2)}} \xi_A^{\mu;\nu} d\Sigma_{\mu\nu} = \\ &-\frac{2}{\kappa}\int_{\Sigma^{(3)}} \xi_{A;\nu}^{\mu;\nu} d\Sigma_\mu = \qquad (3)\\ &2\int_{\Sigma^{(3)}}\left(T_A^\mu - \frac{1}{2}\delta_A^\mu T\right)d\Sigma_\mu.\end{aligned}$$

Comparing (2) with (3) on hyper surface $\mu = 0$, i.e. in observing 3-dimensional space differential of volume $dV^{(3)} = d\Sigma_0$, following relations at $A=0,3$ are obtained.

At $A=0$ the known virial theorem is obtained [8] :

$$\int_{V^{(3)}} T_i^i dV = 0 \quad (i=1,2,3). \qquad (4)$$

At $A=3$ the extension of virial theorem on toroidal systems is obtained. This extension presents the conservation of the momentum angular component :

$$\int_{V^{(3)}} T_3^0 dV = 0. \qquad (5)$$

From the equations of motion (1) one more condition of the stationary CBT existence can be obtained. At $\mu = 0$ from (1) the next equation follows :

$$T^{00}_{;0} + T^{0i}_{;i} = 0.$$

Integrating this zero over 3-volume of system $V^{(3)}$ we obtain in stationary case ( $,_0 = 0$ ) by using Gauss theorem :

$$\oint_{\Sigma^{(2)}=\partial V^{(3)}} T^{0i} d\Sigma_i = 0. \qquad (6)$$

So <u>statement 2</u> is proved : the total flux of particles and field through 2-boundary of stationary CBT $\Sigma^{(2)}$ should be equal to zero.

We assume that the total energy momentum tensor $T^{\mu\nu}$ is the sum of energy momentum tensor of substance $\overset{s}{T}{}^{\mu\nu}$ and energy momentum tensor of field $\overset{f}{T}{}^{\mu\nu}$. Let the charged medium presents an isotropic ideal fluid with energy density $\varepsilon_s$ and hydrodynamic pressure $p_s$. The non-traced electromagnetic field ( $\overset{f}{T}{}^\mu_\mu = 0$ )has energy density $\varepsilon_f$. For the spatial part of the total energy momentum tensor trace we have :

$$T_i^i = \varepsilon_s - 3p_s - \varepsilon_s - \varepsilon_f = -(\varepsilon_f + 3p_s). \qquad (7)$$

Substituting (7) in virial theorem (4) we obtain conditions of stationary CBT existence :

$$E_f = -\int_{V^{(3)}} 3p_s dV, \qquad (8)$$

where $E_f = \int_{V^{(3)}} \varepsilon_f dV$ is the total energy of electromagnetic field in the system volume.

So <u>statement 3</u> follows from (8) : stationary state of compact charged medium (relativistic CBT) in electromagnetic field in vacuum Minkowski space-time is possible a) at negative $\varepsilon_f$ and positive $p_s$, b) at positive $\varepsilon_f$ and negative $p_s$, c) at zero energy of electromagnetic field in the system volume.

If such conditions are not achievable that at given assumptions we can only speak about the long-living quasistationary states of CBT to be used as AMT for controlled thermonuclear fusion.

It is possible that the «negative pressure» effect can result in interaction of CBT with conducting toroidal vacuum vessel walls. This so-called «mirror capture» effect has been observed in charged beams physics [11]. But investigation of this problem is beyond this work.

## 4. Conclusion

General conditions of high power relativistic beam (HPRB) closed configurations to be used for absolute magnetic trap (AMT) formation with the purpose of high temperature long time confinement in limited space region (the problem of control thermonuclear fusion) are considered in this work using the framework of relativistic electrodynamics. In addition, the considered HPRB states can be directly used for obtaining storage elements of low entropic energy.

It is shown that AMT can be formed only by toroidal HPRB own electromagnetic field because only in this case formation of closed magnetic surfaces without singularities is possible. Also the stationary nonradiative[2] states of charged medium in field are possible.

From the system symmetry supposing two time-like and space-like cyclic directions of motion (Killing vectors) presence the virial theorem generalization in case of compact beam torus (CBT) has been received. Also it is shown that CBT stationary state in vacuum is possible only at exotic conditions (for example, in case of negative hydrodynamic pressure inside system). At the non-negative pressure we can speak only about quasistationary long-living states of charged medium in its own and external electromagnetic fields.

It is shown from the common principles (the certain symmetries of compact system presence) by means of virial theorem [8] that the exact plasma balance in its own and external fields does not exist on conditions that the pressure is non-negative in case of compact relativistic insular charged medium. But such plasma balance can approximately exist in "drift approximation" [3,4]. That is we can speak only about the quasistationary states of long time confinement of beam + plasma system. The case of negative pressure is performed to be unusual for the time present. But it can be interesting for obtaining the exact stationary state in combination with the other ways, for example, obtaining the stationary state by means of compact beam torus blowing by external particles flow.

Also the next conclusion of this work can be interesting for applications. The AMT state is strongly unachievable in case of charged medium confinement by means of external field because of singularity presence at magnetic field formation by current windings.

Acknowledgments
We are grateful to Didenko A.N. for his consultation about the «mirror capture» effect and to Lukashin P.Yu. for his help.

---

[2] As magnetic deceleration (synchrotron) radiation of revolving relativistic beam exists strong nonradiative state achievement is problem too.